\documentclass[%
 reprint,
 amsmath,amssymb,
 aps,
]{revtex4-2}

\usepackage{graphicx}
\usepackage{dcolumn}
\usepackage{bm}

\usepackage[T1]{fontenc}
\usepackage{inconsolata}
\usepackage{url}
\usepackage{hyperref}
\usepackage{color}
\usepackage{graphicx}
\usepackage[table,xcdraw]{xcolor}

\usepackage{listings}
\begin{document}

\preprint{APS/123-QED}

\title{The generalized Vaidya spacetime with polytropic equation of state}

\author{Vitalii Vertogradov}
 \altaffiliation[ ]{Physics Department, Herzen State Pedagogical University of Russia, 48 Moika Emb., Saint Petersburg 191186, Russia
SPb branch of SAO RAS, 65 Pulkovskoe Rd, Saint Petersburg 196140, Russia}
 \email{vdvertogradov@gmail.com}

\date{\today}

\begin{abstract}
The process of the gravitational collapse might lead not only to a black hole but also to naked singularity formation. In this paper, we consider the generalized Vaidya spacetime with polytropic and generalized polytropic equations of state. We solve the Einstein and Einstein-Maxwell equations to obtain the explicit form of a mass function. We consider the limiting cases of solutions and find out, that generalized Vaidya spacetime might behave like Vaidya-de Sitter and Bonnor-Vaidya-de sitter solutions. Moreover, we explicitly show, that the part of solution, which depends on the polytropic index, is similar to cosmological fields surrounding both Vaidya and Bonnor-Vaidya black holes. The process of the gravitational collapse has been then considered. We have found out that the conditions of the naked singularity formation don't depend on the polytropic index.\\

{\bf keywords:} Generalized Vaidya spacetime, Gravitational collapse, Polytropic equation of state, Naked singularity, black hole.
\end{abstract}
\maketitle


\section{Introduction}

Nowadays, we have strong experimental evidence of the existence of a black hole through gravitational wave detection~\cite{bib:ligo1, bib:ligo2} and the image of a black hole shadow~\cite{bib:eht1, bib:eht2, bib:eht3}. When a star is at the end of its lifecycle, i.e. when all fuel that supports its stability is exhausted, the star undergoes the continuous gravitational collapse. Depending on the initial profile, this process can lead to white dwarf, neutron stars, or black hole formations. The cosmic censorship conjecture [CCC] states that any singularity must be covered with a horizon. It means that gravitational collapse of sufficient massive star should be inevitable black hole. However, it has been shown that, under physically-relevant conditions, the result of continuous gravitational collapse might be a naked singularity, which is the CCC violation~\cite{bib:joshi, bib:joshi_review}. A thorough investigation of different models revealed that under some physically relevant conditions, the naked singularity might be the result of the gravitational collapse~\cite{bib:vaidya_burst, bib:radial, bib:zek, bib:zek2, bib:dis}. Moreover, the shadow that can cast a naked singularity can't be distinguished from a black hole one~\cite{bib:joshi_shadow}. Naked singularities can be considered as an arena of high energy physics~\cite{bib:joshi_bsw1, bib:joshi_bsw2, bib:joshi_bsw3}.

Oppenheimer and Snyder~\cite{bib:open}  constructed the model of the gravitational collapse with the assumption of the pressureless homogeneous matter. Homogeneous dust collapse always leads to a black hole formation. However, if one considers inhomogeneous dust distribution, then the result of such collapse might be a naked singularity~\cite{bib:homogeneous, bib:homogeneous2}. We say that a naked singularity might form during the gravitational collapse if the following conditions are held:
\begin{itemize}
\item The time of the naked singularity formation is less than the time of the apparent horizon formation;
\item there exists a family of non-spacelike, future-directed geodesics which terminate in the central singularity in the past.
\end{itemize}

Vaidya spacetime~\cite{bib:vaidya}, alternatively referred to as radiating Schwarzschild black hole, is one of the earliest examples of CCC violation~\cite{bib:pap}. The Vaidya spacetime is widely used in many astrophysical applications with strong gravitational fields~\cite{bib:mizner}. In general relativity, Vaidya spacetime assumed to describe the exterior of the radiating stars~\cite{bib:santos1985non}. The pressure at the surface is non-zero, and the star dissipates energy in the form of heat flux. This made it possible to study dissipation and physical features associated with gravitational collapse, as shown by Herrera et al.~\cite{bib:herrera2006some, bib:herrera2012dynamical,bib:joshivaidya}. The backreaction of the accreting matter leads to the vaidya spacetime in spherically-symmetric case~\cite{bib:docu}. The question about the dynamical shadow formation in Vaidya and charged Vaidya spacetimes are discussed in~\cite{bib:germany,bib:japan,bib:ver_shadow}. The horizon structure, entropy and diagonalization  of this solution are investigated for an empty background in~\cite{bib:nelvaidya, bib:nelsurface, bib:kudr, bib:diag, bib:charged_hak}. Vaidya-Kerr like spacetimes and  Vaidya black hole with K-essence and surrounded by cosmological fields are considered in~\cite{bib:kerrv, bib:kerrh, bib:manna1, bib:manna2, bib:manna3, bib:manna4, bib:tur1, bib:tur2, bib:tur3}. Some recent studies of the temperature properties inside the radiating star have been done in~\cite{bib:reddy2015impact, bib:thirukkanesh2012final, bib:thirukkanesh2013}. The Vaidya spacetime can be extended to include both null dust and null string fluids leading to the generalized Vaidya spacetime~\cite{bib:vunk}. A detailed investigation of the properties of the generalized Vaidya spacetime can be found in~\cite{bib:husain1996exact, bib:Radiationstring1998, bib:twofluidatm1999}. The generalized Vaidya spacetime has been used to investigate gravitational collapse ~\cite{bib:Maombi1, bib:Maombi2, bib:ver1, bib:ver2, bib:ver3, bib:myrev, bib:r2}. The conformal symmetries and embedding and other properties are discussed in~\cite{bib:maharaj_conformal, bib:charged_conformal, bib:r4, bib:r1, bib:r3}.  Non-singular black holes are discussed in~\cite{bib:hay, bib:non_singular}. The generalized Vaidya spacetime has the off-diagonal term which may lead to the negative energy states for a particle like in Kerr spacetime~\cite{bib:pen}. The absence of such particles has been proven  in~\cite{bib:verneg}. However, there is a similar effect for the charged particles and a generalized Penrose process may take place in static Reissner-Nordstrom~\cite{bib:rufini} and dynamic Bonnor-Vaidya~\cite{bib:ver_penrose} cases. The forces in Vaidya spacetime are discussed in~\cite{ bib:verforce, bib:verlinear}. Recently, a new generalization of Vaidya spacetime has been found by the gravitational decoupling method~\cite{bib:vernew}, which can describe the Vaidya black hole distorted by dark matter.

The thorough investigation of the gravitational collapse of the generalized Vaidya spacetime revealed the possibility of the naked singularity formation~\cite{bib:Maombi1} for general matter distribution. It is well-known, that the presence of an equation of state introduces a differential relation for a mass function. There has been considered a barotropic~\cite{bib:husain1996exact} and Hagedorn~\cite{bib:vaidya_burst} equations of state. An equation of state is a constitutive relation that provides the link between some state quantities describing the system. Typically, for a collapsing matter an equation of state is provided once the pressure can be expressed as a function of the energy density $P=P(\rho)$. The equations of state that are usually considered for perfect fluids at equilibrium are two. A barotropic relation $P=\alpha \rho$ and polytropic one $P=\alpha \rho^\gamma$. The study of the collapse of a perfect fluid with the polytropic equation of state is more complicated. In both cases, solving the differential equation for the mass function $M(v,r)$ might prove to be unattainable.

The polytropic equation of state has been used in describing the physical properties of stellar objects and anisotropic relativistic stars~\cite{bib:isayev, bib:starsnew, bib:maharaj_stars, bib:musaev, bib:maha}. The generalized Vaidya spacetime with barotropic and polytropic equations of state have been considered in~\cite{bib:husain1996exact}. Despite, the generalized Vaidya spacetime with barotropic equation of state has been widely investigated, the solution with polytropic equation of state, except for short description in~\cite{bib:husain1996exact}, has not been properly investigated yet in scientific literature. In this paper, we consider the simple model of the gravitational collapse with the polytropic equation of state. Motivated by the fact that the Einstein equations for generalized Vaidya spacetime are linear in mass, we have found a new solution of the Einstein equation  which describes the generalized Vaidya spacetime with generalized polytropic equation of state. We also solve the system of the Einstein -Maxwell field equations   with polytropic equation of state for type-II matter field . The solution depends on four parameters (See formulas \eqref{eq:mass1.1}, \eqref{eq:mass2}, \eqref{eq:mass_gen} and \eqref{eq:addtointr} below): two arbitrary functions $C(v)$ and $D(v)$, elictric charge $Q(v)$ and polytropic index $n$. The continuous gravitational collapse depends on these four parameters. We thoroughly investigate the influence of them on the end-state of this process wether it would be naked singularity or black hole. We show that the function $C(v)$ can be associated with a black hole mass and that $D(v)$ and polytropic index $n$ do not have any impact on the end-state of the gravitational collapse. Moreover, we consider several cases of the function $D(v)$ and show that generalized Vaidya spacetime with polytropic equation of state can be Vaidya-de Sitter and Bonnor-Vaidya-de Sitter black holes. The part of solution, which depends on the polytropic index $n$, acts as a cosmological field surrounding both Vaidya and Bonnor-Vaidya black holes.

This paper is organized as follows: in sec II. we consider the exact solution of the Einstein field equations for the generalized Vaidya spacetime with the polytropic equation of state and discuss the gravitational collapse and some limiting cases  of the solution. In sec. III. we consider the system of the Einstein-Maxwell equations and find the analytical solution. In sec. IV we solve the Einstein field equation and obtain a new solution which describes the generalized Vaidya spacetime with generalized polytropic equation of state. Sec. V is the conclusion. 

Throughout the paper, we will use the geometrized system of units in which $G=1=c$, also we use the signature $-+++$. We also introduce the following notations:

\begin{eqnarray}
M'&\equiv &\frac{\partial M}{\partial r},\nonumber \\
\dot{M}&\equiv &\frac{\partial M}{\partial v}.
\end{eqnarray}

\section{Generalized Vaidya solution with polytropic equation of state}

The Vaidya spacetime can be extended to include both null dust and null string fluids leading to the generalized Vaidya spacetime~\cite{bib:vunk}. It has the form
\begin{equation} \label{eq:metric}
ds^2=-\left(1-\frac{2M(v,r)}{r}\right)dv^2+2\varepsilon dvdr+r^2d\Omega^2,
\end{equation}
where $M(v,r)$ is the mass function which depends on both the areal coordinate $r$ and Eddington time $v$, $\varepsilon=\pm1$ depending on ingoing or outgoing radiation. $d\Omega^2=d\theta^2+\sin^2\theta d\varphi^2$ is the metric on unit two-sphere. Our main concern is in gravitational collapse. Hence, we put $\varepsilon=+1$. If the mass function $M(v,r)$ depends only on the Eddington time $v$ then the metric \eqref{eq:metric} reduces to the Vaidya spacetime with energy-momentum tensor in the form of the null dust
\begin{equation} \label{eq:emtvaidya}
T^{(ND)}_{ik}=\mu(v,r)l_il_k,
\end{equation}
here $l_i=\delta^0_i$ is the nul vector ($l^il_i=0$) and $\mu(v,r)$ is the energy dencity of the null dust
\begin{equation} \label{eq:nul_dust}
\mu(v,r)=\frac{2\dot{M}(v)}{r^2} \,.
\end{equation}
The generalized Vaidya spacetime \eqref{eq:metric} is supported with the energy-momentum tensor which represents the combination of the null dust \eqref{eq:emtvaidya} and the null strings
\begin{equation} \label{eq:emtgeneralized}
T^{(NS)}_{ik}=(\rho+P)(l_in_k+l_kn_i)+Pg_{ik},
\end{equation}
where $n^i$ is another null vector with properties
\begin{equation}
n^in_i=0,~~ n^il_i=-1,
\end{equation}
$\rho$ and $P$ are the energy density and pressure respectively. They are given by
\begin{eqnarray} \label{eq:density}
\rho(v,r)&=&\frac{2M'(v,r)}{r^2},\nonumber \\
P(v,r)&=&-\frac{M''(v,r)}{r}.
\end{eqnarray}
The null vectors $l^i$ and $n^i$ have the following form
\begin{eqnarray}
l_{i}&=&\delta^0_i,\nonumber \\
n_{i}&=&\frac{1}{2} \left (1-\frac{2M}{r} \right )\delta^0_{i}-\delta^1_{i}.
\end{eqnarray}
Thus, the total energy-momentum tensor for the generalized Vaidya spacetime \eqref{eq:metric} is given by
\begin{equation} \label{eq:emt}
T_{ik}^{total}=T_{ik}^{(ND)}+T_{ik}^{(NS)} \,.
\end{equation}
We are interested in the end-state of the continuous gravitational collapse . If the end-state is a naked singularity then the energy-momentum tensor \eqref{eq:emt} should be physically-relevant, i.e. it should satisfy the energy conditions~\cite{bib:pois}. For generalized Vaidya spacetime they read~\cite{bib:Maombi1}:
\begin{itemize}
\item Weak and strong energy conditions.
\begin{equation} \label{eq:wec}
\mu>0,~~ P\geq 0,~~ \rho \geq 0,~~ \mu \neq 0.
\end{equation}
\item Dominant energy condition.
\begin{equation} \label{eq:dec}
\mu>0,~~ \rho\geq P \geq 0,~~ \mu \neq 0.
\end{equation}
\end{itemize}
The equation of the state, which is usually considered for perfect fluids at equilibrium, is polytropic one
\begin{equation} \label{eq:state}
P=\alpha \rho^\gamma \,.
\end{equation}
Where $\alpha$ is constant and another constant $\gamma$ is usually written as
\begin{equation}
\gamma=1+\frac{1}{n} \,.
\end{equation}
Here $n$ is a polytropic index of the system. If $\gamma=1$ one comes to the linear equations of the state. This solution has been widely investigated in scientific literature and it is known as Husain solution~\cite{bib:husain1996exact}. In our consideration, we assume that $\gamma \neq 1$. The solution with $\gamma\neq 1$ has been also obtained in the paper~\cite{bib:husain1996exact}, but its properties have not been studied in detail so far. It can be also shown that the condition $n\leq 5$ should be held, otherwise the matter cloud has no boundary at equilibrium~\cite{bib:joshi_review}. 

The Einstein equations for the generalized Vaidya spacetime \eqref{eq:metric} with the matter distribution of the form \eqref{eq:emt} with the polytropic equation of the state \eqref{eq:state} lead to the following differential equation
\begin{equation} \label{eq:dif_equation1}
M''+2^\gamma \alpha \frac{(M')^\gamma}{r^{2\gamma-1}}=0 \,.
\end{equation}
In order to solve this equation we introduce a new function $W=M'$\footnote{One should note that we can make this replacement only with the made assumption $\gamma\neq 1$.} then \eqref{eq:dif_equation1} becomes
\begin{eqnarray} \label{eq:solution1}
&&W'+2^\gamma \alpha \frac{W^\gamma}{r^{2\gamma -1}}=0 \rightarrow\nonumber \\
&&W(v,r)=\left(D(v)-2^{\gamma-1}\alpha r^{2-2\gamma}\right)^{\frac{1}{1-\gamma}}.
\end{eqnarray}
Where $D(v)$ is an integration function of the time $v$. 
From \eqref{eq:solution1} one can easily obtain the mass function in the form
\begin{equation} \label{eq:full_solution1}
M(v,r)=\int Wdr+C(v),
\end{equation}
where $C(v)$ is another function of integration.
To consider the gravitational collapse, one should evaluate the integral $\int Wdr$ in \eqref{eq:full_solution1}. We consider separately three cases:
\begin{itemize}
\item $D(v)\equiv 0$;
\item $D(v) is small$;
\item The case of the most general $D(v)$.
\end{itemize}

\subsection{The case $D(v)\equiv 0$}

We begin our consideration with the most simple case when an integration constant $D(v)$ is zero. In this case, the function $W(v,r)$ reads
\begin{equation}
W(v,r)=\left(-1\right)^{n}\frac{\alpha^{\xi}r^2}{2},
\end{equation}
where $\xi =\frac{1}{1-\gamma}$. The integration in \eqref{eq:full_solution1} gives
\begin{equation}
\int Wdr=\left(-\right)^n\frac{\alpha^\xi r^3}{6} \,.
\end{equation}
The mass function is then given by
\begin{equation} \label{eq:mass1.1}
M(v,r)=C(v)+\left(-\right)^n\frac{\alpha^\xi r^3}{6} \,,
\end{equation}
and the spacetime \eqref{eq:metric} takes the following form
\begin{equation} \label{eq:metric1.1}
ds^2=-\left(1-\frac{2C(v)}{r}-\left(-1\right)^n\frac{\alpha^\xi r^2}{3}\right)dv^2+2dvdr+r^2d\Omega^2 \,.
\end{equation}
This spacetime is the Vaidya-de Sitter spacetime~\cite{bib:vaidya_sitter}, where $C(v)$ is associated with Vaidya mass function $M(v)$ and $\left(-1\right)^n\alpha^\xi$ is interpreted as cosmological constant $\Lambda$. Note, that depending on the sign $\alpha$ and value of a polytropic index $n$, we obtain either Vaidya-de Sitter spacetime $\Lambda>0$ or Vaidya-anti-de Sitter spacetime $\Lambda<0$.
As we mentioned in the introduction, the naked singularity is the result of the gravitational collapse if the following conditions are held:
\begin{itemize}
\item The time of the singularity formation is less than the time of the apparent horizon formation;
\item there is a family of non-spacelike, future-directed geodesics which terminate in the central singularity in the past.
\end{itemize}
The apparent horizon equation for the spacetime \eqref{eq:metric1.1} has the form
\begin{equation} \label{eq:ah1.1}
1-\frac{2C(v)}{r}-\left(-1\right)^n\frac{\alpha^\xi r^2}{3}=0 \,.
\end{equation}
At the time of the naked singularity formation $v=0$, the apparent horizon is absent if the limit
\begin{equation}
\lim\limits_{v\rightarrow 0, r\rightarrow 0} \frac{2C(v)}{r}=X_1 \,,
\end{equation}
exists and finite. 
For considering the existence of a family of non-spacelike, future-directed geodesic, we consider a family of radial null geodesics
\begin{equation} \label{eq:geodesic1.1}
\frac{dv}{dr}=\frac{2}{1-\frac{2C(v)}{r}-\left(-1\right)^n\frac{\alpha^\xi r^2}{3}} \,.
\end{equation}
Here such geodesics terminate in the central singularity in the past and future-directed if the limit
\begin{equation} \label{eq:limit1.1}
\lim\limits_{v\rightarrow0, r\rightarrow 0} \frac{dv}{dr}=X_0 >0 \,,
\end{equation}
finite. Note, that for the gravitational collapse case, $X_0=0$ doesn't suit us because $v=const.$ is the ingoing null geodesic and we are interested only in the outgoing one. 
We consider the function $C(v)$ in the linear form, i.e. 
\begin{equation}
C(v)=\mu v,~~ \mu>0.
\end{equation}
Now, by taking the limit \eqref{eq:limit1.1}, the geodesic equation \eqref{eq:geodesic1.1} becomes
\begin{eqnarray}
X_0&=&\frac{2}{1-2\mu X_0} \rightarrow\nonumber \\
0&=&2\mu X_0^2-X_0+2 \rightarrow\nonumber \\
X_0^{\pm}&=&\frac{1}{4\mu}\left(1\pm \sqrt{1-16\mu}\right),
\end{eqnarray}
and we obtain a well-known result: the naked singularity is the result of the gravitational collapse if the condition $\mu \leq \frac{1}{16}$ is held. This result was obtained in the paper~\cite{bib:joshivaidya}. The gravitational collapse of Vaidya-de Sitter spacetime has been also considered in~\cite{bib:vah}.  Note, that the outcome of the gravitational collapse doesn't depend on the sign of $\alpha$.

\subsection{The case of small $D(v)$ }

In the previous subsection, we have found out that the function $C(v)$ can be associated with a black hole mass $M(v)$. Here, we will find the influence of the function $D(v)$ on the gravitational collapse if it is negligible small during the gravitational process. For this purpose, we introduce a dimensionless parameter $\delta$
\begin{equation}
D(v)=\delta d(v),~~ \delta\ll 1,
\end{equation}
and expand $W(v,r)$ with respect to this parameter. The function $W(v,r)$ \eqref{eq:solution1} reads
\begin{equation} \label{eq:w2.1}
W\approx \left( -1 \right)^{-n}\frac{\alpha^{-n} r^2}{2}-\frac{n\delta d(v) }{2^{\frac{n+1}{n}}\alpha^{n+1}} r^{2\frac{n+1}{n}} \,.
\end{equation}
The integration gives
\begin{equation}
\int Wdr \approx \left(-1\right) ^{-n}\frac{\alpha^{-n} r^3}{6}-\beta(v)r^{2\gamma+1},
\end{equation}
where
\begin{equation}
\beta(v)=\frac{ n\delta d(v)}{(2\gamma+1)2^\gamma \alpha^{n+1}}  \,.
\end{equation}
The mass function then takes the form
\begin{equation} \label{eq:mass2.1}
M(v,r)=C(v)+\left(-1\right) ^{-n}\frac{\alpha^{-n} r^3}{6} -\beta(v)r^{2\gamma+1} \,.
\end{equation}
and the generalized Vaidya spacetime then reads
\begin{equation} \label{eq:metric2.1}
ds^2=-\left(1-\frac{2C(v)}{r}-\left(-1\right) ^{-n}\frac{\alpha^{-n} }{3}r^2+2\beta(v)r^{2\gamma}\right)dv^2+2dvdr +r^2d\Omega^2 \,.
\end{equation}
Where $C(v)$ and $\left(-1\right) ^{-n}\alpha^{-n} $ have the same interpretation of the mass $M(v)$ and the cosmological constant $\Lambda$. The fourth term in $g_{00}$ can be interpreted as another cosmological field. For example, if $\gamma=\frac{3}{2}$ then the term $\beta(v) r^3$ can be associated with phantom field~\cite{bib:menase} surrounding a Vaidya black hole~\cite{bib:tur1}. 

We should also mention that if $\dot{\beta}<0$ then we satisfy weak energy condition. However, if $\dot{\beta}>0$ then there is a region near singularity
\begin{equation}
r\leq \left(\frac{2\dot{C}(v)}{\dot{\beta}(v)}\right)^{\frac{1}{2\gamma+1}} \,,
\end{equation}
where the weak energy condition is violated. This problem is the same as in Bonnor-Vaidya spacetime~\cite{bib:energy_violation} where there is also the region where weak energy condition is violated.

The apparent horizon equation gives
\begin{equation}
1-\frac{2C(v)}{r}-\left(-1\right)^n\frac{\alpha^n}{3}r^3+2\beta(v)r^{2\gamma}=0\,.
\end{equation}
Like in the previous case, for the existence of the naked singularity, one must demand that the limit
\begin{equation}
\lim\limits_{v\rightarrow 0, r\rightarrow0}\frac{C(v)}{r}=X_1 \,,
\end{equation}
should be finite. 

The radial null geodesic for the metric \eqref{eq:metric2.1} reads
\begin{equation} \label{eq:geodesic2.1}
\frac{dv}{dr}=\frac{2}{1-\frac{2C(v)}{r}-\left(-1\right)^n\frac{\alpha^n}{3}r^2+2\beta(v)r^{2\gamma}} \,.
\end{equation}

We should impose the condition on $\beta(v)$, i.e. it should be  finite at the time $v=0$ i.e. $\beta(0)=\beta_0$. Then, assuming the linear mass function
\begin{equation}
C(v)=\mu v,~~ \mu>0 \,,
\end{equation}
and taking the limit of \eqref{eq:geodesic2.1} by using the deffinition \eqref{eq:limit1.1}, one obtains
\begin{eqnarray}
X_0&=&\frac{2}{1-2\mu X_0} \rightarrow\nonumber \\
X_0^{\pm}&=&\frac{1}{4\mu}\left(1\pm \sqrt{1-16\mu}\right),
\end{eqnarray}
i.e. the same condition for the naked singularity occurence $\mu \leq \frac{1}{16}$ like in the previous case.

\subsection{The general case}

We have considered some particular cases of the first integration function $D(v)$ so far. Here, we assume that $D(v)$ is in the most general form. then the generalized Vaidya spacetime \eqref{eq:metric} takes the form
\begin{equation} \label{eq:metric3.1}
ds^2=-\left(1-\frac{2C(v)}{r}-\frac{2}{r}\int W(v,r)dr\right)dv^2+2dvdr+r^2d\Omega^2 \,.
\end{equation}
The apparent horizon equation then gives
\begin{equation}
1-\frac{2C(v)}{r}-\frac{2}{r}\int W(v,r)dr=0 \,,
\end{equation}
and the radial null geodesics reads
\begin{equation} \label{eq:geodesic3.1}
\frac{dv}{dr}=\frac{2}{1-\frac{2C(v)}{r}-\frac{2}{r}\int W(v,r)dr} \,.
\end{equation}
From the condition $M(0,0)$~\cite{bib:Maombi1}, we obtain that $D(0)=0$. Hense, we can expand $M(v,r)$ near the singular point $(0,0)$
\begin{equation}
M(v,r)\approx M'_0r+\dot{M}_0v \,.
\end{equation}
Where we made the following notations
\begin{eqnarray}
\lim\limits_{v\rightarrow 0, r\rightarrow 0}\frac{dM}{dr}=M'_0,\nonumber \\
\lim\limits_{v\rightarrow0, r\rightarrow 0}\frac{dM}{dv}=\dot{M}_0.
\end{eqnarray}
From the fact that $\gamma>1$ and solution \eqref{eq:solution1} we can conclude that
\begin{equation}
\lim\limits_{v\rightarrow 0, r\rightarrow 0}\frac{d}{dr} \left(\int W(v,r)dr\right) =0 \,.
\end{equation}
So, like in previous cases, assuming $C(v)=\mu v$ and considering the geodesic equation \eqref{eq:geodesic3.1} with definition \eqref{eq:limit1.1}, we obtain the same condition for the naked singularity formation $\mu \leq \frac{1}{16}$.

The integral $\int W(v,r)dr$ can be, in general, interpreted as some cosmological field surrounding the Vaidya black hole. 

\section{Polytropic equation of state and Maxwell field equations}

In this section, we will find the solution of the Einstein field equations coupling with electromagnetic field. We, like in the previous section, assume the spacetime in the form \eqref{eq:metric}
\begin{equation} \label{eq:metric2}
ds^2=-\left(1-\frac{2M(v,r)}{r}\right)dv^2+2dvdr+r^2d\Omega^2 \,.
\end{equation}
However, the energy-momentum tensor \eqref{eq:emt} we assume to be
\begin{equation} \label{eq:emt2}
T_{ik}^{total}=T_{ik}^{(ND)}+T_{ik}^{(NS)}+E_{ik} \,.
\end{equation}
Here, as earlier, the first two terms are null dust \eqref{eq:emtvaidya} and null strings \eqref{eq:emtgeneralized}. The correction $E_{ik}$ 
\begin{equation} \label{eq:max}
E_{ik}=\frac{1}{4\pi}\left(F_{ij}F^j_k-\frac{1}{4}g_{ik}F_{jl}F^{jl}\right),
\end{equation}
is an electromagnetic contribution. $F_{ik}$ is antisymmetric electromagnetic field tensor. It obeys the Maxwell equations
\begin{eqnarray} \label{eq:max_eq}
F_{ik,l}&+&F_{li,k}+F_{kl,i}=0,\nonumber \\
F^{ik}_{;k}&=&-4\pi J^i.
\end{eqnarray}
Without any loss of generality, the electromagnetic vector potential can be chosen as~\cite{bib:bonor, bib:lake}
\begin{equation}
A_i=\frac{Q(v)}{r}\delta^0_i \,.
\end{equation}
Where $Q(v)$ is associated with a black hole dynamical electric charge. From \eqref{eq:max_eq} one obtains non-vanishing component of the electromagnetic tensor $F_{ik}$
\begin{equation} \label{eq:emft}
F_{10}=-F_{01}=\frac{Q(v)}{r^2} \,.
\end{equation}
And the electromagnetic contribution $E_{ik}$ to the total energy-momentum tensor is
\begin{equation}
E_{ik}=\frac{Q^2}{r^2}\eta_{ik},
\end{equation}
where $\eta_{ik}$ is the components of flat spacetime metric tensor.

The Einstein field equations with energy-momentum tensor \eqref{eq:emt2} for the metric \eqref{eq:metric2} are given by
\begin{eqnarray} \label{eq:einstein}
\mu(v,r)&=&\frac{2}{r^2}\dot{M},\nonumber \\
\rho(v,r)&=&\frac{2}{r^2}M'-\frac{Q^2}{r^4},\nonumber \\
P(v,r)&=&-\frac{1}{r}M''-\frac{Q^2}{r^4}.
\end{eqnarray}
Imposing the polytropic equation of state on two last equations \eqref{eq:einstein}  and using the definitions $P$ and $\rho$ \eqref{eq:density}, we obtain
\begin{equation} \label{eq:einstein_eq}
\frac{1}{r}M''+\frac{Q^2}{r^2}+\alpha\left(\frac{2}{r^2}M'-\frac{Q^2}{r^2}\right)^\gamma=0 \,.
\end{equation}
We introduce the function $W(v,r)$ by relation
\begin{equation} \label{eq:w_def}
W(v,r)=M'-\frac{Q^2}{2r^2},
\end{equation}
and substitute it into the Einstein field equation \eqref{eq:einstein_eq} in order to obtain the followin expression
\begin{equation}
\frac{1}{r}W'+\alpha \left(\frac{2W}{r^2}\right)^\gamma =0\,.
\end{equation}
The solution of this equation is
\begin{equation} \label{eq:w2}
W(v,r)=\left(D(v)-2^{\frac{1}{n}}\alpha r^{2-2\gamma}\right)^{\frac{1}{1-\gamma}} \,.
\end{equation}
In order to obtain the mass function $M(v,r)$, one should substitute \eqref{eq:w_def} into \eqref{eq:w2}. After integration, one obtains
\begin{equation} \label{eq:mass2}
M(v,r)=\int Wdr-\frac{Q^2}{2r}+C(v) \,.
\end{equation}
Note, that the integral $\int Wdr$ is the same as in the uncharged case. In the previous section, we found out that it might be interpreted as a cosmological field and it doesn't affect the condition for the naked singularity occurence. Here, the situation is the same. Thus, we don't need to consider all three cases of general $D(v)$, negligible small or zero $D(v)$. Instead, we will consider the simplest case $D(v)\equiv 0$ and find out the conditions for the naked singularity formation. These conditions will be the same in all other cases. 

With the mass function of the form \eqref{eq:mass2}, the generalized Vaidya spacetime \eqref{eq:metric2} takes the form
\begin{equation} \label{eq:genmet}
ds^2=-\left(1-\frac{2C(v)}{r}+\frac{Q^2}{r^2}-\frac{2}{r}\int Wdr \right)dv^2+2dvdr+r^2d\Omega^2 \,.
\end{equation}
Here we assume $D(v)\equiv 0$. Then this spacetime reads
\begin{equation} \label{eq:metric_final}
ds^2=-\left(1-\frac{2C(v)}{r}+\frac{Q^2}{r^2}-\left(-1\right)^{-n}\frac{\alpha^{-n}}{3}r^2\right)dv^2+2dvdr+r^2d\Omega^2 \,.
\end{equation}
This spacetime is the charged Vaidya-(anti) de Sitter spacetime~\cite{bib:tur2}. Let's denote $M(v)\equiv C(v)$ and $\Lambda=\left(-1\right)^{-n}\alpha^{-n}$. In these notations, the apparent horizon equation reads
\begin{equation}
1-\frac{2M(v)}{r}+\frac{Q^2(v)}{r^2}-\frac{\Lambda}{3}r^2=0\,.
\end{equation}
This spacetime, as we mentioned above, has the region 
\begin{equation} \label{eq:ah2}
r\leq \frac{Q\dot{Q}}{\dot{M}} \,, 
\end{equation}
where the weak energy condition is violated. However, particles can't get into this region due to the repulsive Lorentz force~\cite{bib:no_violation}.

This spacetime, under some values of $Q\,, M\,, \Lambda$ might have the eternal naked singularity formation\footnote{Under notion 'eternal' we mean that the singularity might form during the gravitational collapse and will never be covered with the apparent horizon} and only one cosmological horizon. 
The radial null geodesic has the form
\begin{equation} \label{eq:geodesic1.2}
\frac{dv}{dr}=\frac{2}{1-\frac{2M(v)}{r}+\frac{Q^2}{r^2}-\frac{\Lambda}{3}r^2} \,.
\end{equation}
If we assume the mass and charge functions of the form
\begin{equation}
M(v)\equiv \mu v,~~ Q^2=\nu^2 v^2 \,.
\end{equation}
and taking into account the deffinition \eqref{eq:limit1.1}, we arrive at
\begin{eqnarray} \label{eq:naked}
X_0&=&\frac{2}{1-2\mu X_0+\nu^2 X_0^2} \rightarrow\nonumber \\
\nu^2 X_0^3&-&2\mu X_0^2+X_0-2=0.
\end{eqnarray}
The last equation is the cubic one. It means that in general, it admits at least one real root. Note that the last equation \eqref{eq:naked} is negative at $X_0=0$ and positive at $X_0=2+\frac{2\mu}{\nu^2}$ i.e. the last equation has the positive finite root in the interval $X_0^{r} \in \left(0\,, \frac{2\mu}{\nu^2}\right)$. Hence, the limit is positive and finite and the naked singularity might form during the gravitational collapse. This result coincides with the one obtained in~\cite{bib:charged_collapse1}. Note that regardless of the value $\gamma$, the naked singularity formation will depend only on conditions obtained in \eqref{eq:naked} because the integral \eqref{eq:mass2} doesn't influence the condition of the naked singularity occurrence. 

\section{The generalized polytropic equation of state}

In this section, we consider generalized Vaidya spacetime with the equation of state
\begin{equation} \label{eq:genstate}
P=k\rho +\alpha \rho^\gamma \,.
\end{equation}
If one assume $\alpha =0$, then \eqref{eq:genstate} becomes a barotropic equation of state $P=k\rho$, which is widely investigated in literature~\cite{bib:husain1996exact}. Here, we briefly remind the basic properties:
\begin{itemize}
\item $k=1$ corresponds to the stiff fluid and solution is the charged Vaidya spacetime;
\item $k=\frac{1}{3}$ corresponds to the radiation;
\item $k=0$ is the equation of state describing dust;
\item $k=-\frac{2}{3}$ corresponds to the quentesence;
\item $k=-1$ is the vaccuum energy and
\item $k=-\frac{4}{3}$ corresponds to phantom field.
\end{itemize}
The pure polytropic part $P=\alpha \rho^\gamma$ may be, due to Bose-Einstein condensates with repulsive ($\alpha >0$) or attractive ($\alpha<0$) self-interaction, or have another origin~\cite{bib:polytropic_universe}. 

The Einstein equations for the generalized Vaidya spacetime \eqref{eq:metric} with energy density $\rho$ and pressure $P$ defined in \eqref{eq:density} with the generalized polytropic equation of state \eqref{eq:genstate} lead to the following second-order differential equation:
\begin{equation} \label{eq:einstein_general}
-\frac{M''}{r}=k\left(\frac{2M'}{r^2}\right)+\alpha \left(\frac{2M'}{r^2}\right)^\gamma \,.
\end{equation}

We again introduce the function $W$ by relation
\begin{equation} \label{eq:wgen}
W(v,r)\equiv \frac{2M'}{r^2} \,.
\end{equation}
Then the Einstein equation \eqref{eq:einstein_general}, in terms of new function $W$, reads as
\begin{equation} \label{eq:dif_gen}
-W'r=(2k+2)W+2\alpha W^\gamma \,.
\end{equation}
To solve this equation one needs to consider two cases.

\subsection{The case $k\neq -1$}

In this case, the solution of the differential equation \eqref{eq:dif_gen} is

\begin{equation} \label{eq:wgen_sol}
W=\frac{(2k+2)^{n}D(v) r^{-(2k+2)}}{\left( 2k+2-2\alpha D(v)r^{-(2k+2)(\gamma-1)}\right)^{-\xi}} \,.
\end{equation}
Now, we should substitute the function $W$ \eqref{eq:wgen} into \eqref{eq:wgen_sol} and the integration gives the mass function in the form

\begin{equation} \label{eq:mass_gen}
M(v,r)=C(v)+\frac{1}{2}\int \frac{(2k+2)^{-\xi}D(v) r^{-2k}}{\left( 2k+2-2\alpha D(v)r^{-(2k+2)(\gamma-1)}\right)^{-\xi}} dr \,.
\end{equation}

Note that in the case of $\alpha =0$ i.e. one considers only barotropic equation of state $P=k\rho$, the mass function $M(v,r)$ becomes
\begin{equation} \label{eq:husain}
M(v,r)=C(v)+D(v)r^{1-2k} \,,
\end{equation}
which corresponds to the Husain solution~\cite{bib:husain1996exact, bib:ver2}.

Here we need to evaluate the integral in the right hand-side of the equation \eqref{eq:mass_gen}. Note, if $D(v)\equiv 0$ then the mass function 
\begin{equation}
M(v,r)=C(v) \text{for} D(v)\equiv 0,
\end{equation}
is the Vaidya mass function. If we consider the small integration function $D(v)\ll 1$ then the mass function $M(v,r)$ reduces to the Husain solution \eqref{eq:husain}. The gravitational collapse, in this case, can lead to the naked singularity formation~\cite{bib:ver2}.

\subsection{The case $k=-1$}

This case, despite the much simpler (in comparison with the previous subsection) differential equation, leads to the solution which can be hardly  analysed analytecaly. The solution of the differential equation \eqref{eq:dif_gen} is given by
\begin{equation}
W(v,r)=\left(D(v)+\ln r^{-2\alpha (1-\gamma)} \right)^\xi \,.
\end{equation}
And we can formally write the solution in the form
\begin{equation} \label{eq:addtointr}
M(v,r)=C(v)+\int W(v,r)dr \,.
\end{equation}
However, this solution can't be analysed within this paper.

\section{Conclusion}

In this paper, we have obtained the three solutions of the Einstein field equations. The first one describes the generalized Vaidya spacetime with the type-II matter field satisfying the polytropic equation of the state. The mass function of a solution has the general form \eqref{eq:full_solution1}. Except for short note in~\cite{bib:husain1996exact}, this solution has not been properly investigated in scientific literature so far. In order to understand what the end-state of the gravitational collapse might be, one should consider the integral $\int Wdr$. In the case of uncharged spacetime, we have considered three cases. The first one leads to the Vaidya-de Sitter spacetime, the second one leads to Vaidya spacetime surrounded by the combination of the cosmological fields. In both cases, there is not influence of the integral $\int Wdr$ on the naked singularity formation. In general case when the integration function $D(v)$ is supposed to be general, we have concluded that this integral $\int Wdr$ may act as a cosmological field. 

By considering the generalized Vaidya spacetime coupling with the electromagnetic field, the solution has the form \eqref{eq:mass2}. In the limiting case $D(v)=0$ the resulting spacetime is charged Vaidya-de Sitter one. Again, considering the gravitational collapse to find the conditions of the naked singularity occurrence, we have found out that these conditions don't depend upon the integral $\int Wdr$, i.e. the polytropic index $\gamma$ doesn't have any impact on these conditions. 

The third solution corresponds to the generalized equation of state \eqref{eq:genstate} and it is a new solution. The solution for zero or negligible small integration function $D(v)$ is either the Vaidya spacetime or Husain solution respectively. 

The main result of this paper is that the end-state of the continuous gravitational collapse, despite more complex equation of state, is still defined by only the mass function $M(v)$ (Vaidya case) or both mass $M(v)$ and charge $Q(v)$ functions (Bonnor-Vaidya case). 

There is a really important essue regarding the strength of the naked singularity. It is important to show that the naked singularity is gravitationally strong, otherwise it does not violate CCC. However, we have shown, that polytropic index does not affect the end-state of the gravitational collapse. It means that the conditions for the naked singularity to be gravitationally strong are the same like in Vaidya and Bonnor-Vaidya models. It was shown that singularities in these models are gravitationally strong.

The generalized Vaidya spacetime with the polytropic equation of state, under some physically-relevant conditions,  violates the cosmic censorship hypothesis. However, the singularities appear in a wide variety of forms and it seems at present not feasible to rule out all singularities with only a single theorem. If, however, cosmic censorship holds, it should be in a highly refined and fine-tuned form only. The naked singularity formation, on the other hand, might play an important role in astrophysics because such ultra-dense regions can tell us about the effects of quantum gravity which are possible and might be visible to the faraway observer.  They can be also considered as a source of the gamma ray bursts~\cite{bib:joshi_burst, bib:vaidya_burst}. 

\textbf{acknowledgments}: The work was performed as part of the SAO RAS government contract approved by the Ministry of Science and Higher Education of the Russian Federation.

\end{document}